\documentstyle[twoside,fleqn,espcrc2,fps]{article}

\newcommand{\AmS}{{\protect\the\textfont2
  A\kern-.1667em\lower.5ex\hbox{M}\kern-.125emS}}

\def\lsi{\raise0.3ex\hbox{$<$\kern-0.75em\raise-1.1ex\hbox{$\sim$}}}
\def\gsi{\raise0.3ex\hbox{$>$\kern-0.75em\raise-1.1ex\hbox{$\sim$}}}

\newcommand{\R}{{\kern+.25em\sf{R}\kern-.78em\sf{I} 
  \kern+.78em\kern-.25em}}
\newcommand{\C}{{\kern+.25em\sf{C}\kern-.50em\sf{I} \kern+.50em\kern-.25em}}

\title{Fast evaluation and locality of overlap fermions}

\author{W. Bietenholz
\address{ Institut f\"{u}r Physik, Humboldt Universit\"{a}t zu Berlin,
Invalidenstr. 110, D-10115 Berlin, Germany \\
$^{{\rm b}}$ Fachbereich Physik, Universit\"{a}t Wuppertal,
D-42097 Wuppertal, Germany},
I. Hip $^{{\rm b}}$ and K. Schilling $^{{\rm b}}$
\thanks{Talk presented by W.B. at LATTICE 2001.}
}
\begin{document}

\begin{abstract}

In order to construct improved overlap fermions,
we start from a short ranged approximate Ginsparg-Wilson
fermion and insert it into the overlap formula.
We show that its polynomial evaluation
is accelerated considerably compared to the standard
Neuberger fermion. In addition the degree of locality is 
strongly improved.

\vspace*{-5mm}

\end{abstract}

\maketitle

\section{IMPROVED OVERLAP FERMIONS}

The Ginsparg-Wilson relation (GWR) for a lattice Dirac operator $D$ 
reads \cite{GW}
\begin{equation}
\{ D , \gamma_{5} \} = 2 D R \gamma_{5} D \ ,
\ (R {\rm ~ local},~ \{ R,\gamma_{5}\} \neq 0).
\end{equation}
We choose $R_{x,y} = \frac{1}{2\mu} 
\delta_{x,y}$, $(\mu > 0)$, and $D^{\dagger} = \gamma_{5}D\gamma_{5}$,
hence the GWR simplifies to
\begin{equation} \label{gwr}
A^{\dagger}A = \mu^{2} \ , \quad A:= D - \mu \ .
\end{equation}
It guarantees an exact --- though lattice modified ---
chiral symmetry in even dimensions \cite{ML}, and parity symmetry 
in odd dimensions \cite{odd}. 
In both cases, the anomaly arises from the measure.

Given some Dirac operator $D_{0}$
(local, no doublers), we can enforce the GWR on
$A_{0}=D_{0}-\mu$ by the overlap formula
$A_{ov}=\mu A_{0}/ \sqrt{A_{0}^{\dagger}A_{0}}$, so that
$D_{ov} = A_{ov} + \mu$ obeys the GWR (for $\mu$ in some allowed
interval).
The standard Neuberger fermion uses the Wilson operator
$D_{0}= D_{W}$ \cite{HN}.

The guide-line for our concept to {\em improve} overlap fermions \cite{WB}
is the following observation: if $D_{0}$ is already a GW fermion
(with respect to a fixed kernel $R$), then $D_{ov}=D_{0}$.
We can explicitly construct an approximate solution to the GWR,
use it as $D_{0}$, and then we expect that the overlap formula only 
causes a small ``chiral correction'',
\begin{equation} \label{chicor}
D_{ov} \approx D_{0} \ .
\end{equation}
Therefore, we first construct a short-ranged
approximate GW fermion, which includes elements of a truncated
perfect action, and then we insert it as $D_{0}$ into the overlap
formula. The resulting $D_{ov}$ has an exact chiral symmetry,
and we expect a high level of {\em locality}, good scaling and approximate
rotation invariance, because these properties hold for $D_{0}$, and
we rely on relation (\ref{chicor}). Moreover, we predict a {\em fast
convergence} under iterative evaluation of $D_{ov}$, because we start
off in the right vicinity.

All these properties have been tested and confirmed in the 2-flavour 
Schwinger model \cite{BH}. Here we report on convergence and locality
in quenched QCD.

\vspace*{-2mm}

\section{THE HYPERCUBE FERMION (HF)}

Our construction of a short-ranged approximate GW fermion is based
on the perfect free fermion \cite{BW}, which is truncated to a unit 
hypercube.
It consists of a vector term and a scalar term,
$D(x,y) = \rho_{\mu}(x-y)\gamma_{\mu} + \lambda (x-y)$,
where $\rho_{\mu}(x-y)$, $\lambda (x-y) \neq 0$ only for
$\vert x_{i} - y_{i} \vert \leq 1$, $(i=1\dots 4)$.
The couplings are given in Ref.\ \cite{BBCW}.
The first step is a ``minimal gauging'' of this HF,
which means that the free couplings are attached to the shortest
lattice paths only. 
This formulation provides already an excellent rotation behaviour,
but it suffers from strong additive mass renormalization \cite{BBCW}.

To cure this problem, we attach a ``link amplification factor''
$1/u$ to each link and tune it to $u_{crit} < 1$ to approach
at the chiral limit. Using different factors for the vector term 
and the scalar term further improves the GW approximation .

Next we add a fat link, i.e. each link variable
is substituted by a linear combination: $(1-\alpha )$ ``direct link''
+ $\alpha /6$ $\sum$ ``staple terms''. 
Finally we also include a clover term.

With this modest set of parameters we optimize the QCD spectrum
at $\beta =6$. In Ref.\ \cite{lat2000} we show the full spectrum 
for typical configurations on a $4^{4}$ lattice, as well as the 
eigenvalues with the least real parts on a $8^{4}$ lattice.

\section{POLYNOMIAL OVERLAP EVALUATION AND LOCALITY}

We consider two ways to evaluate the overlap Dirac operator
by means of polynomial expansions, which have appeared in the
literature.\\
1) We introduce the Hermitean operator $H_{0} \doteq \gamma_{5}A_{0}$,
and evaluate 
\begin{equation}
D_{ov} = \mu \Big( 1 + \gamma_{5} \frac{H_{0}}{\sqrt{H_{0}^{2}}} \Big) \ .
\end{equation}
The last term corresponds to a sign function $\epsilon (x) = 1$ $(-1)$ 
for $x>0$ $(x<0)$, which is approximated by a polynomial.\\
2) We simply expand $1/\sqrt{x}$ by a polynomial, and insert
$A_{0}^{\dagger}A_{0}$ for $x$.

For the sign expansion, the convergence in the degree of the
polynomial is exponential, if $ x \in [-1,1]$.
Hence we first re-scale $H_{0} \to \bar H_{0}$, so that the spectrum
$\sigma (\bar H_{0}) \subset [-1,1]$.
What matters is the density of eigenvalues (EVs) of $\bar H_{0}$
around 0, where the polynomial converges most painfully due to the
discontinuity. For the HF, $\sigma (H_{0})$ has two peaks around
$\pm 1$, which are shifted slightly towards the origin for $\bar H_{0}$,
leaving typically (at $\beta =6$) a wide gap without EVs
of $\bar H_{0}$ between about $[-0.3, 0.3]$. On the other hand, 
$H_{0,W}$ has a broad spectrum, ranging from about $[-6,6]$,
and after re-scaling one does obtains EVs near 0.
Histograms are shown in Ref.\ \cite{lat2000}.

For the $1/\sqrt{x}$ expansion the situation is similar: here we re-scale so
that $\bar x \in [\delta , 1]$, $(\delta >0)$. It is crucial that
$\delta$ (the lower bound of the re-scaled spectrum) does not become very 
small; getting close to the singularity slows down the convergence.
Again, the Wilson spectrum is very broad to start with, and re-scaling
usually leads to small $\delta$, whereas it is kept larger for the HF.

\begin{figure}[hbt]
\def\fpsangle{270}
\epsfxsize=45mm
\fpsbox{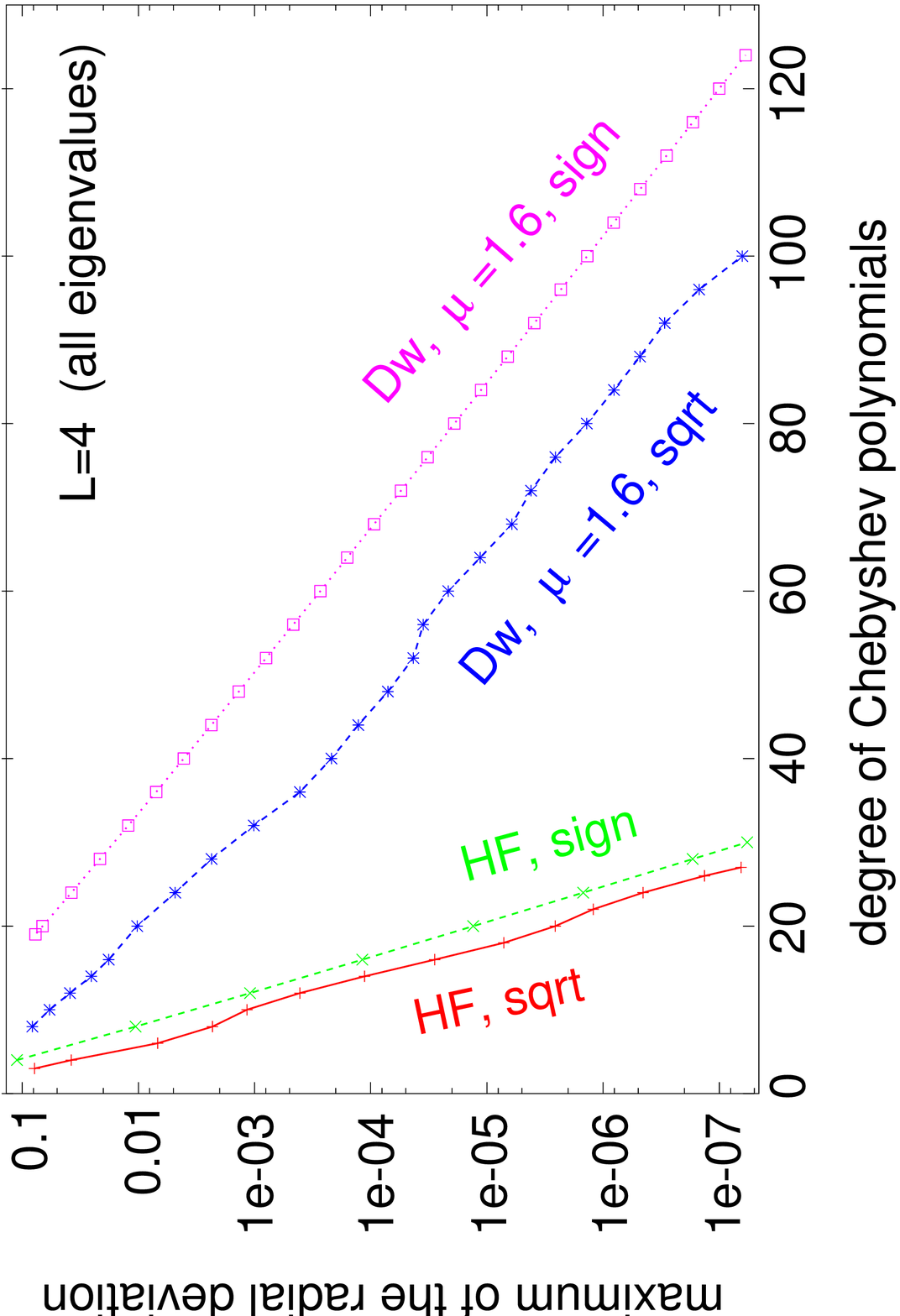}
\epsfxsize=46mm
\fpsbox{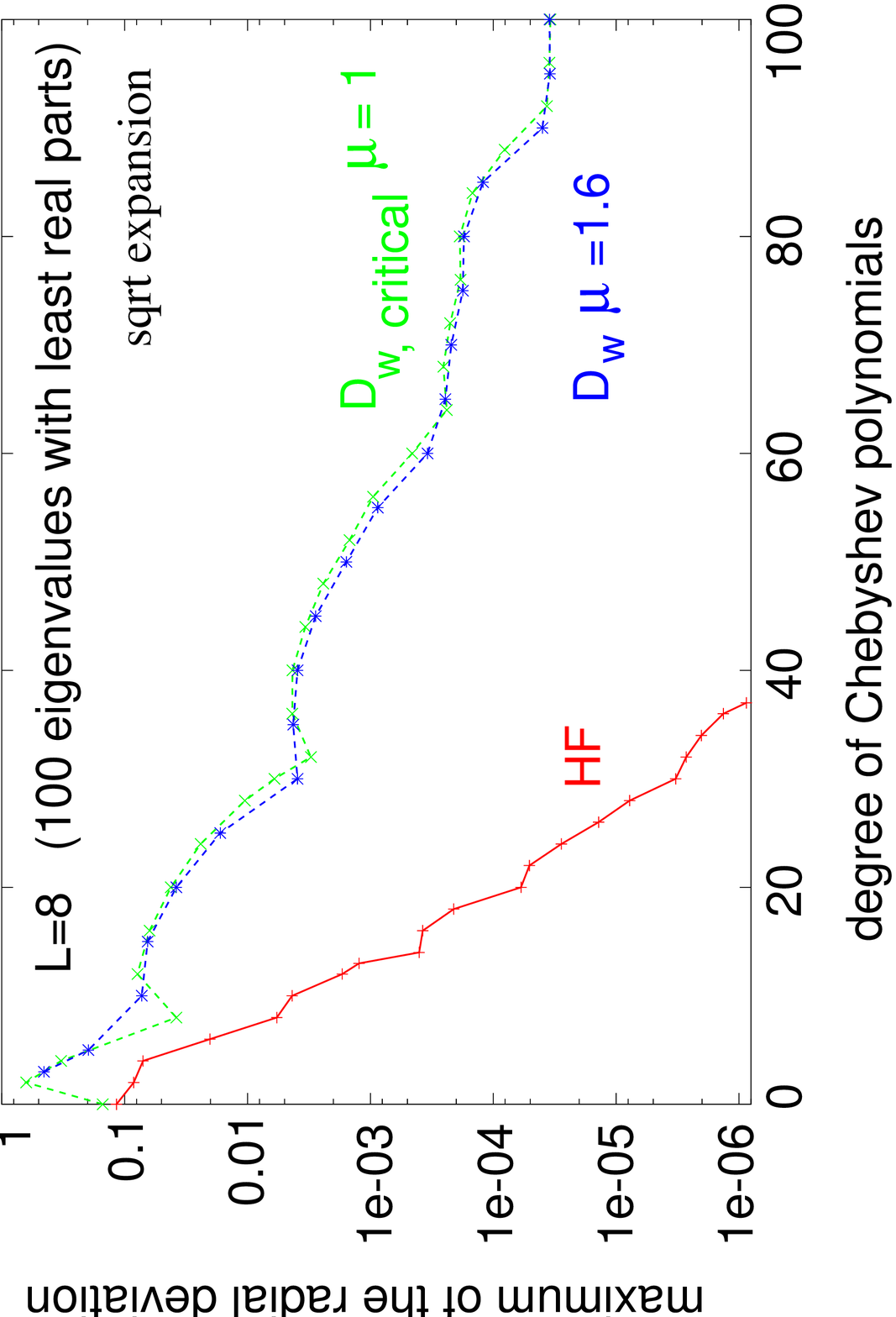}
\vspace{-7mm}
\caption{\it{The maximal deviation of an energy eigenvalue
of approximate overlap Dirac operators (after re-scaling)
from the unit circle in $\C$.}}
\vspace{-8mm}
\label{fig-conv}
\end{figure}

In Fig.\ \ref{fig-conv} we illustrate the convergence rate 
in QCD at $\beta =6$ using Chebyshev polynomials.
As an example, we discuss the $\epsilon (x)$ expansion on the $4^{4}$
lattice. For $D_{0} = D_{W}$, the maximal (mean) deviation of the
EVs of $\frac{1}{\mu}D$ from the unit circle in $\C$ (with center 1)
behaves typically as $d^{max}_{W}=\exp (-0.134n)$
($d^{mean}_{W}=0.13 \exp (-0.134n$)), where $n$ is the degree of the 
Chebyshev polynomials. The corresponding decays for $D_{0}=D_{HF}$ 
amount to $d^{max}_{HF}=\exp (-0.737n)$ ($d^{mean}_{HF}=0.1 \exp(-0.737n)$).

If we fix some affordable degree $n$, the precision of the GW
approximation is superior for the HF by a factor 
$d_{W}^{max}/d^{max}_{HF} = \exp (0.6n)$, which takes a very
considerable magnitude for realistic numbers like $n=20 \dots 100$.

On the other hand, we could fix a certain accuracy
$d^{max}$ which we consider necessary, so the required degree
$n$ compares as $n_{W} / n_{HF} = 5.5$.
Referring to this number, the computational overhead of the HF, 
which amounts to a factor $\leq 20$, is compensated in part.
It now takes only a relatively small progress in the scaling
behaviour to make up for the remaining overhead.

\begin{figure}[hbt]
\def\fpsangle{270}
\epsfxsize=50mm
\fpsbox{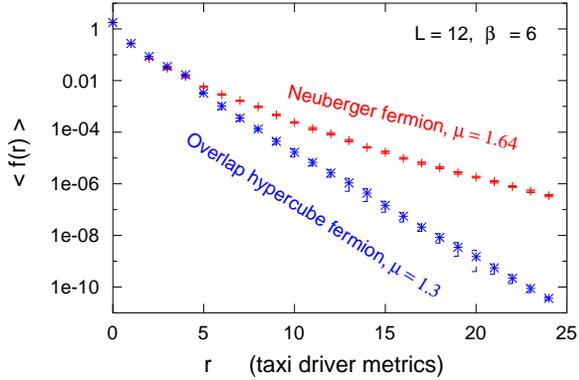}
\vspace{-8mm}
\caption{\it{Comparison of the degree of
locality for different overlap fermions.}}
\vspace{-10mm}
\label{fig-loc}
\end{figure}

Fig.\ \ref{fig-loc} shows that 
locality is clearly improved for the overlap-HF
compared to the Neuberger fermion. Following Ref.\ \cite{HJL},
we measure the ``maximal correlation'' $f(r)$ over a distance $r$.

\begin{figure}[hbt]
\def\fpsangle{270}
\epsfxsize=41mm
\fpsbox{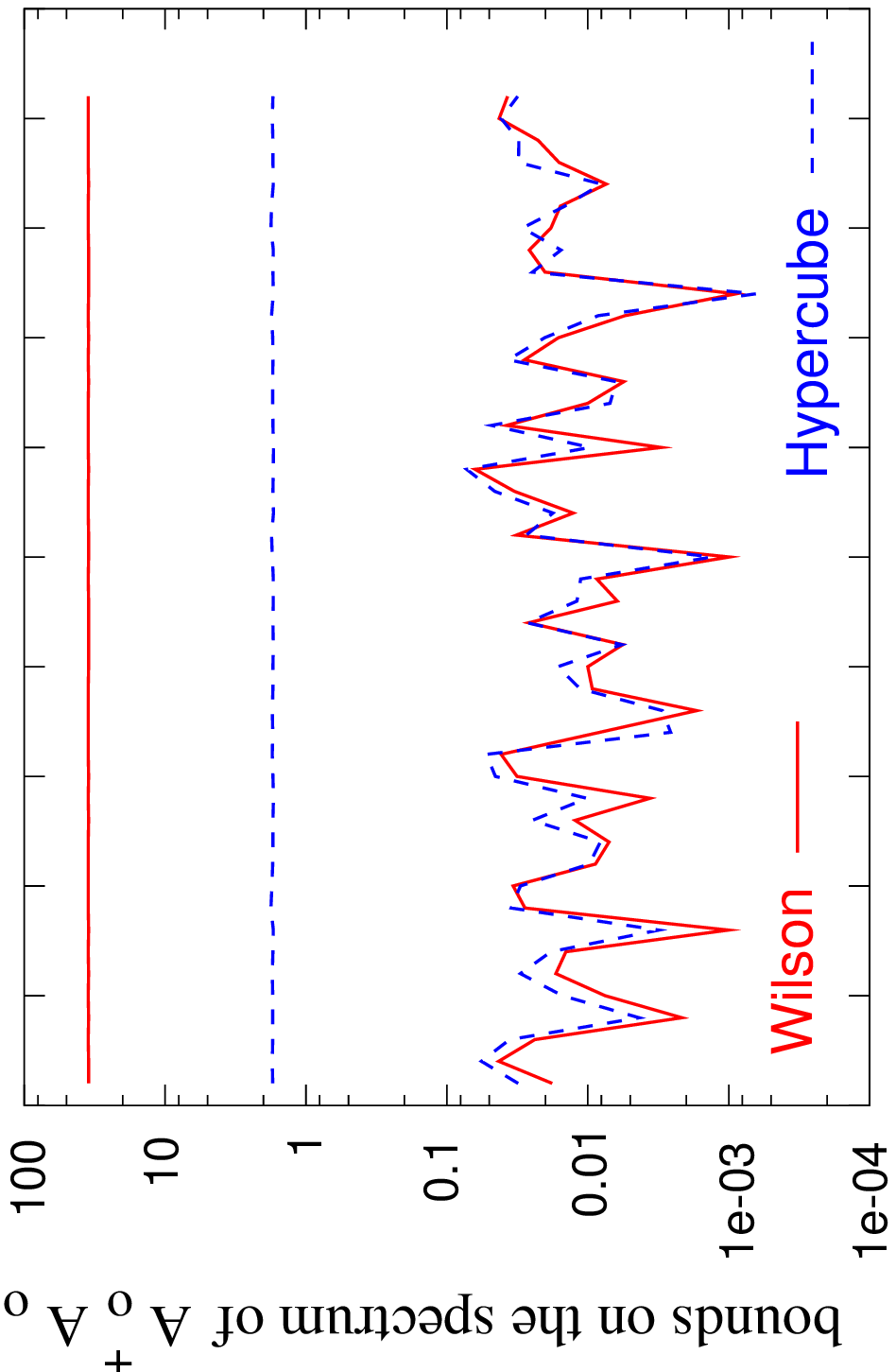}
\epsfxsize=46mm
\fpsbox{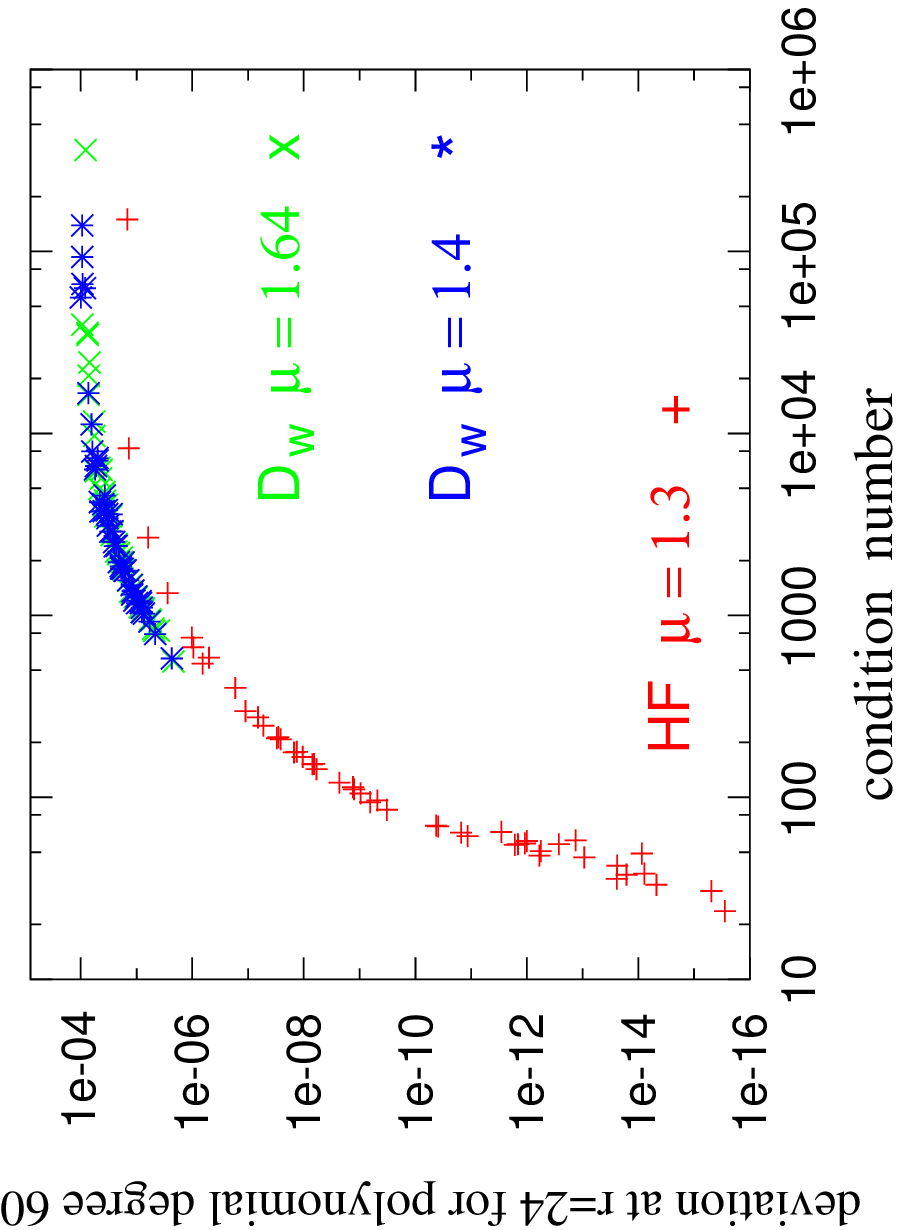}
\vspace{-10mm}
\caption{\it{On top:\ history of the spectral bounds. 
Their ratio ( = condition number) 
is essential for the polynomial
convergence rate (see below).}}
\vspace{-11mm}
\label{fig-cond}
\end{figure}

We stay on the $12^{4}$ lattice and we observe also here
a clear progress in the speed of convergence. This is related to
the condition number $c$, i.e.\ the ratio of the upper and
the lower bound of the spectrum of $A_{0}^{\dagger}A_{0}$.
Fig.\ \ref{fig-cond} (on top) shows that the improved
condition number for the HF is essentially due to the {\em decrease of
the upper bound}. To illustrate the effect of $c$ on the convergence 
rate, Fig.\ \ref{fig-cond} (below) shows the deviation of 
$f(r=24)$ from the precise result
if we use a moderate polynomial of degree $n=60$.
This deviation depends smoothly on $c$, up to an extra gap
which makes the deviation for $D_{0} = D_{W}$ yet a bit worse than 
$c$ suggests, even at the optimal parameter $\mu =1.64$.

\vspace*{-5mm}

\section{CONCLUDING REMARKS}

\vspace*{-1mm}

Compared to the standard Neuberger fermion, the overlap-HF gains
significantly in the convergence rate under polynomial evaluation.
We have studied the evolution of the full operator in the polynomial
at $\beta =6$, which might be roughly comparable to the situation
at stronger coupling, say $\beta =5.75$, after treating the O(10) worst
modes separately, as it is often done.

Moreover, we observed a superior degree of locality.
This suggests that the overlap formula is applicable
up to stronger coupling (the limit for $\beta$ was discussed in 
Ref.\ \cite{lat2000}).

The scaling quality remains to be tested; we are currently
measuring meson dispersions. Another important question is the
applicability of preconditioning techniques. They were
already applied successfully to the HF without overlap \cite{SESAM}.


\end{document}